\documentclass[aps,prd,nofootinbib,twocolumn,showpacs]{revtex4-1}
\usepackage{graphicx}
\usepackage{amssymb}
\usepackage{epstopdf}
\usepackage{hyperref}

\usepackage{amsmath}
\usepackage{bbold}
\usepackage{color}

\begin{document}

%\title{Evading the off-shell Higgs width constraint with new light scalar resonances}
\title{Hiding a Higgs width enhancement from off-shell $gg \ (\to h^*) \to ZZ$ measurements}

\author{Heather E.~Logan}
\email{logan@physics.carleton.ca} 
\affiliation{Ottawa-Carleton Institute for Physics, Carleton University, 1125 Colonel By Drive, Ottawa, Ontario K1S 5B6, Canada}

\date{December 24, 2014}                                  % Activate to display a given date or no date

\begin{abstract}
\noindent
Measurements of the off-shell Higgs boson production cross section in $gg \ (\to h^*) \to ZZ$ have recently been used by the CMS and ATLAS collaborations to indirectly constrain the total width of the Higgs boson.  I point out that the interpretation of these measurements as a Higgs width constraint can be invalidated if additional neutral Higgs boson(s) are present with masses below about 350~GeV.
\end{abstract}


\maketitle 

%%%%%%%%%%%%%%%%%%%%%%%%%%%%%%%%%%%%%%%%%%%%%%
\section{Introduction}

The measurement of the properties of the recently-discovered~\cite{Aad:2012tfa} 125~GeV Higgs boson is central to particle physics over the next decade~\cite{Brock:2014tja}.  These properties, in particular the couplings of the Higgs boson to other Standard Model (SM) particles, probe the underlying cause of electroweak symmetry breaking.  Searches for exotic decays of the Higgs boson further provide a probe for new physics which may be coupled to the SM solely through Higgs interactions~\cite{Patt:2006fw}.  Sensitivity to the Higgs boson couplings at the CERN Large Hadron Collider (LHC) comes primarily from measurements of ``signal strengths,'' i.e., of the rates of Higgs production and decay in particular production modes and into particular final states.  On the Higgs resonance, the couplings can be parameterized by a collection of multiplicative factors $\kappa_i$~\cite{LHCHiggsCrossSectionWorkingGroup:2012nn} that modify the corresponding SM couplings.  The on-resonance rate in a particular production and decay channel can then be expressed in the narrow-width approximation as
\begin{equation}
	{\rm Rate}_{ij} = \sigma_i \frac{\Gamma_j}{\Gamma_{\rm tot}}
	= \kappa_i^2 \sigma_i^{\rm SM} \frac{\kappa_j^2 \Gamma_j^{\rm SM}}
	{\sum_k \kappa_k^2 \Gamma_k^{\rm SM} + \Gamma_{\rm new}},
\end{equation}
where $\sigma_i$ is the Higgs production cross section in production mode $i$, $\Gamma_j$ is the Higgs decay partial width into final state $j$, $\Gamma_{\rm tot}$ is the total width of the Higgs boson, the corresponding quantities in the SM are denoted with a superscript, and $\Gamma_{\rm new}$ represents the partial width of the Higgs boson into new, non-SM final states.

Rate measurements in all accessible production and decay channels are combined in a fit to extract the coupling factors $\kappa_i$.  This fit possesses a well-known flat direction~\cite{Zeppenfeld:2000td}; for example, one can imagine a scenario in which all the coupling modification factors have a common value $\kappa_i \equiv \kappa > 1$ and there is a new, unobserved contribution to the Higgs total width, $\Gamma_{\rm new} > 0$.  In this case the Higgs production and decay rates measurable at the LHC are given by
\begin{equation}
	{\rm Rate}_{ij} = \frac{\kappa^4 \sigma_i^{\rm SM} \Gamma_j^{\rm SM}}{\kappa^2 \Gamma_{\rm tot}^{\rm SM} + \Gamma_{\rm new}}.
\end{equation}
All measured Higgs production and decay rates will be equal to their SM values if
\begin{equation}
	\kappa^2 = \frac{1}{1 - {\rm BR}_{\rm new}},
\end{equation}
where the Higgs branching ratio into nonstandard final states is
\begin{equation}
	{\rm BR}_{\rm new} \equiv \frac{\Gamma_{\rm new}}{\Gamma_{\rm tot}}
	= \frac{\Gamma_{\rm new}}{\kappa^2 \Gamma_{\rm tot}^{\rm SM} + \Gamma_{\rm new}}.
\end{equation}
In particular, a simultaneous enhancement of all the Higgs couplings to SM particles can mask, and be masked by, the presence of new decay modes of the Higgs (such as light jets~\cite{Berger:2002vs}) that are not directly detected at the LHC.\footnote{Measuring such an enhancement in the Higgs couplings would be straightforward at a lepton-collider Higgs factory such as the International Linear Collider (ILC), where a direct measurement of the total Higgs production cross section in $e^+e^- \to Zh$ can be made with no reference to the Higgs decay branching ratios by using the recoil mass method (see, e.g., Ref.~\cite{Baer:2013cma}).}

This flat direction can be bounded by imposing additional theoretical assumptions, such as the absence of new, unobserved Higgs decay modes~\cite{Zeppenfeld:2000td,Lafaye:2009vr} or the imposition of $\kappa_W, \kappa_Z \leq 1$, which is valid when the Higgs sector contains only isospin doublets and/or singlets~\cite{Duhrssen:2004cv}.  However, viable models exist in which $\kappa_{W,Z}$ can be significantly larger than $1$, such as the Georgi-Machacek model with isospin-triplet scalars~\cite{Georgi:1985nv,Chanowitz:1985ug} (for a recent update of the allowed enhancement of $\kappa_{W,Z}$ see Ref.~\cite{Hartling:2014aga}), extensions of the Georgi-Machacek model with larger isospin representations~\cite{Logan:2015xpa}, and a model with an isospin-septet scalar mixing with the usual doublet~\cite{Kanemura:2013mc,Hisano:2013sn}.  In these models it is straightforward to simultaneously enhance the Higgs couplings to vector bosons and to fermions, thereby reopening the flat direction.

Naturally, the elimination of this loophole in LHC Higgs coupling measurements has become a high priority.
Two novel techniques have been proposed since the Higgs boson discovery that offer direct sensitivity to the product of the Higgs production and decay couplings in selected channels, and hence, via the corresponding signal strength, to the Higgs total width.  The first makes use of the tiny shift in the reconstructed Higgs resonance position in the $gg \to h \to \gamma\gamma$ invariant mass spectrum caused by interference between the signal and the continuum background~\cite{Martin:2012xc,Martin:2013ula,Dixon:2013haa}.  This method is robust against new-physics effects, but is not very sensitive: with the full 3000~fb$^{-1}$ high-luminosity LHC data-set this technique may ultimately be able to constrain $\Gamma_{\rm tot} < 15 \, \Gamma_{\rm tot}^{\rm SM}$~\cite{Dixon:2013haa}.
The second uses the contribution of off-shell $gg \to h^* \to ZZ$ production to the total $gg \to ZZ$ rate above the $ZZ$ production threshold~\cite{Kauer:2012hd,Caola:2013yja,Campbell:2013una}.\footnote{The $WW$ final state can provide additional sensitivity~\cite{Campbell:2011cu,Campbell:2013wga}.}  Away from the $h$ resonance, the off-shell $gg \to h^* \to ZZ$ cross section is proportional to $\kappa_g^2 \kappa_Z^2$, while on resonance the corresponding $gg \to h \to ZZ$ cross section is proportional to $\kappa_g^2 \kappa_Z^2/\Gamma_{\rm tot}$.  Thus a combination of the on- and off-resonance measurements can be used to place an indirect constraint on $\Gamma_{\rm tot}$.

This second technique has already been used by the CMS and ATLAS experiments to set upper bounds on the Higgs total width~\cite{Khachatryan:2014iha,Aad:2015xua}:
\begin{eqnarray}
	\Gamma_{\rm tot} &\leq& 5.4 \, \Gamma_{\rm tot}^{\rm SM} \qquad \qquad \ {\rm (CMS)}, \nonumber \\
	\Gamma_{\rm tot} &\leq& (4.8 - 7.7) \, \Gamma_{\rm tot}^{\rm SM} \quad {\rm (ATLAS)},
\end{eqnarray}
where the range in the quoted ATLAS limit represents theoretical uncertainty in the background cross section.  Along the SM-mimicking flat direction $\kappa_i^2 \equiv \kappa^2 = 1/(1 - {\rm BR}_{\rm new})$, these bounds translate into quite stringent bounds on the common coupling modification factor $\kappa$ and the Higgs branching ratio to non-SM final states:
\begin{eqnarray}
	|\kappa| &\leq& 1.52, \qquad \quad \quad {\rm BR}_{\rm new} \leq 0.60, \nonumber \\
	|\kappa| &\leq& 1.48 - 1.67, \quad {\rm BR}_{\rm new} \leq 0.54 - 0.64,
\label{eq:limits}
\end{eqnarray}
for CMS and ATLAS, respectively.

It has already been pointed out~\cite{Gainer:2014hha,Englert:2014aca,Cacciapaglia:2014rla,Azatov:2014jga,Englert:2014ffa,Buschmann:2014sia} that these bounds must be interpreted with great caution.  In particular, the CMS and ATLAS bounds rely on the assumption that the product of coupling modification factors $\kappa_g^2 \kappa_Z^2$ is independent of the center-of-mass energy $\sqrt{\hat s} \equiv m_{ZZ}$ of the process.  This assumption comes into play because the on-resonance signal strength measurements depend on the coupling values at $m_{ZZ} = 125$~GeV, while the sensitivity to the off-shell Higgs contribution to continuum $gg \to ZZ$ comes from the high $ZZ$ invariant mass region $m_{ZZ} > 2 m_t \simeq 350$~GeV.  This assumption can break down if the $ggh$ coupling is modified due to a new weak-scale particle running in the loop~\cite{Englert:2014aca,Englert:2014ffa} or if either of the couplings is modified due to the contribution of momentum-dependent dimension-six operators~\cite{Gainer:2014hha,Englert:2014aca,Cacciapaglia:2014rla,Azatov:2014jga,Englert:2014ffa,Buschmann:2014sia} that parameterize the effects of new physics at a scale above the direct kinematic reach of the measurement.

In this paper I point out another way in which an enhancement of the Higgs total width can go undetected in the $gg \ (\to h^*) \to ZZ$ analysis.  I consider the scenario in which the Higgs couplings to top quarks and $W,Z$ bosons are modified due to an extended Higgs sector, and there are no new light colored degrees of freedom running in the $ggh$ loop, so that $\kappa_g = \kappa_t$ neglecting light quark contributions.  When the product of coupling modification factors $\kappa_t \kappa_Z \neq 1$, the discovered Higgs boson $h$ by itself no longer unitarizes the $t \bar t \to Z Z$ scattering amplitude at high energy~\cite{Chanowitz:1978uj,Appelquist:1987cf,Bhattacharyya:2012tj} (indeed, the resulting linear growth of this amplitude with increasing center-of-mass energy is the origin of the sensitivity to $\kappa_g \kappa_Z \neq 1$ of the $gg \, (\to h^*) \to ZZ$ cross section measurement at high $m_{ZZ}$~\cite{Englert:2014aca}).  Of course, in a renormalizable model, unitarity is restored once contributions from the additional Higgs boson(s) are included.  If the new Higgs boson(s) are light compared to the $m_{ZZ}$ range in which the LHC $gg \ (\to h^*) \to ZZ$ measurement obtains its sensitivity, their contribution to the Higgs-exchange amplitude largely cancels the modification due to $\kappa_t \kappa_Z > 1$.
In this way the presence of new Higgs boson(s) below about 350~GeV can render the off-shell $gg \ (\to h^*) \to ZZ$ analysis insensitive to an enhancement of the $h$ couplings, and hence to a corresponding new unobserved decay width of $h$.

In the next section I give the details of the calculation, and I conclude in Sec.~\ref{sec:conclusions}.

%%%%%%%%%%%%%%%%%%%%%%%%%%%%%%%%%%%%%%%%%%%%%%
\section{The effect of a second scalar resonance}
\label{sec:gf}

For simplicity I consider the situation in which a single additional (undiscovered) neutral Higgs boson $H$ completes the unitarization of $t \bar t \to ZZ$.  In CP-conserving two Higgs doublet models~\cite{Branco:2011iw} this is the second CP-even neutral Higgs boson; in the Georgi-Machacek model~\cite{Georgi:1985nv,Chanowitz:1985ug} this is the second custodial-singlet Higgs boson.  $H$ is usually taken to be heavier than $h$, though this is not necessary for what follows.  Unitarity of the $t \bar t \to Z Z$ scattering amplitude requires that
\begin{equation}
	\kappa_t^h \kappa_Z^h + \kappa_t^H \kappa_Z^H = 1,
\end{equation}
where the couplings of $h$ and $H$ relative to those of the SM Higgs are distinguished by a superscript.\footnote{The couplings of $h$ and $H$ to top quarks and $W$ and $Z$ bosons can be written in the generic form
\begin{eqnarray}
	\kappa_t^h &=& \frac{\cos\alpha}{\cos \theta_H}, \qquad 
		\kappa_{W,Z}^h = \cos\alpha \cos\theta_H - A \sin\alpha \sin\theta_H, \nonumber \\
	\kappa_t^H &=& \frac{\sin\alpha}{\cos\theta_H}, \qquad
		\kappa_{W,Z}^H = \sin\alpha \cos\theta_H + A \cos\alpha \sin\theta_H,
\end{eqnarray}
where $\alpha$ is the mixing angle that diagonalizes the $h$--$H$ mass-squared matrix and $\cos\theta_H \equiv v_{\phi}/v$, with $v = (\sqrt{2} G_F)^{-1/2}$ being the SM Higgs vacuum expectation value and $v_{\phi}$ being the vacuum expectation value of the doublet that gives rise to the top quark mass.  The coefficient $A$ depends on the isospin of the additional scalar multiplet: $A = 0$ for the SM Higgs mixed with a singlet scalar, $A = 1$ for a two Higgs doublet model, $A = \sqrt{8/3}$ for the Georgi-Machacek model, and $A = 4$ for the SM Higgs mixed with a scalar septet.  When $A > 1$, one can have $\kappa_{W,Z}^h > 1$ for appropriately-chosen values of $\alpha$ and $\theta_H$.  The situation $\kappa_t^h \kappa_Z^h > 1$ is not possible in the singlet scalar extension of the SM.}  In the enhanced-coupling scenario we can then write
\begin{eqnarray}
	\kappa_t^h \kappa_Z^h &\equiv& 1 + \Delta > 1, \nonumber \\
	\kappa_t^H \kappa_Z^H &=& - \Delta.
\end{eqnarray}

Away from the $h$ and $H$ resonances, the sum of the amplitudes for $gg \to h^* \to ZZ$ and $gg \to H^* \to ZZ$, normalized to the corresponding SM amplitude for a Higgs of mass $m_h$, reduces to
\begin{equation}
	\frac{\mathcal{M}_h + \mathcal{M}_H}{\mathcal{M}_{h_{\rm SM}}} \equiv \mathcal{\bar M} 
	= (1 + \Delta) - \Delta \frac{p^2 - m_h^2}{p^2 - m_H^2},
\label{eq:fullamp}
\end{equation}
where $p^2 = m_{ZZ}^2$ is the square of the invariant mass of the final-state $ZZ$ system.

\begin{figure}
\resizebox{0.5\textwidth}{!}{\includegraphics{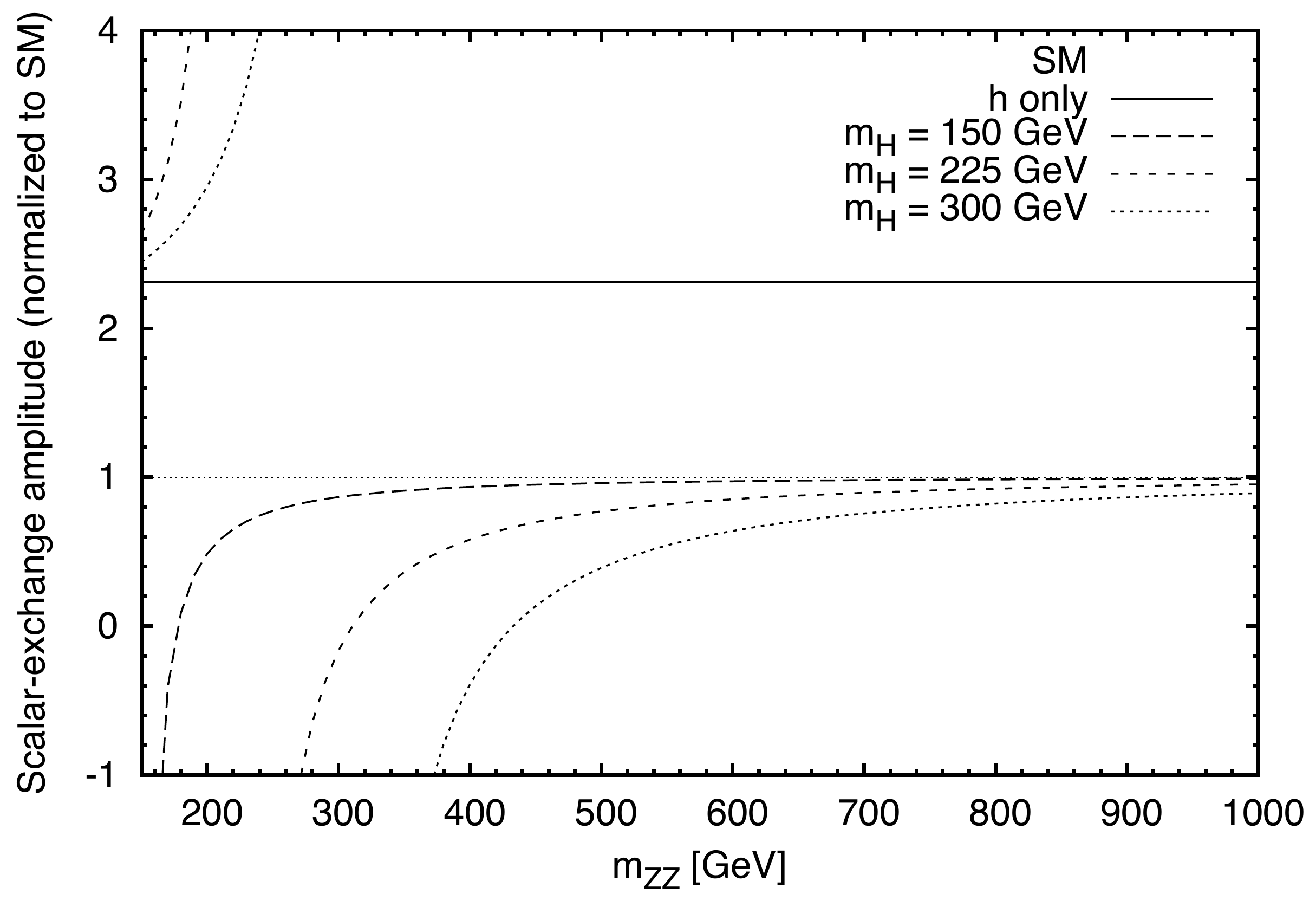}}
\caption{Normalized scalar amplitude $\mathcal{\bar M}$ as a function of the $ZZ$ invariant mass, for the SM, $h$ exchange only, and $h$ and $H$ exchange for three sample values of the $H$ mass.  I take $\kappa_g^h \kappa_Z^h \equiv 1 + \Delta = 2.31$ corresponding to the CMS limit given in Eq.~(\ref{eq:limits}).}
\label{fig}
\end{figure}

When $p^2 \gg m_h^2, m_H^2$, this becomes,
\begin{equation}
	\mathcal{\bar M} = 1 - \Delta \frac{(m_H^2 - m_h^2)}{p^2} 
	+ \mathcal{O} \left( \Delta \frac{m_H^4}{p^4} \right).
\end{equation}
In particular, the scalar-exchange amplitude rapidly approaches its SM value $\mathcal{\bar M} = 1$ with increasing $p^2$.\footnote{Reference~\cite{Englert:2014ffa} studied the scenario in which the SM Higgs boson mixes with a light isospin-singlet scalar, leading to $\kappa_t \kappa_Z < 1$ for the SM-like state, and observed the same return to the SM amplitude in the high-$p^2$ limit.}  This is shown in Fig.~\ref{fig}, where I plot $\mathcal{\bar M}$ for the product of couplings $\kappa_g^h \kappa_Z^h \equiv 1 + \Delta = 2.31$ corresponding to the CMS limit given in Eq.~(\ref{eq:limits}), and $m_H = 150$, 225, and 300~GeV (dashed curves).\footnote{When $m_h^2 < m_H^2 < p^2$, the amplitude in Eq.~(\ref{eq:fullamp}) is in fact \emph{suppressed} compared to the SM expectation.  This is due to the proximity of the $H$ resonance with its negative product of couplings $\kappa_t^H \kappa_Z^H = -\Delta$, which interferes destructively with the $h$ exchange diagram.  This mirrors the \emph{enhancement} of the $4\ell$ differential cross section at low invariant masses, shown in Fig.~8 of Ref.~\cite{Englert:2014ffa}, caused by the high-energy tail of the $H$ resonance in the case of a light isospin singlet mixed with the SM Higgs, in which case the product of couplings $\kappa_t^H \kappa_Z^H = \cos^2\chi > 0$.}  These should be compared to the situation in which only $h$ exchange is considered, in which $\mathcal{\bar M} = 1 + \Delta$ (solid horizontal line), in particular in the region $m_{ZZ} > 350$~GeV where the LHC off-shell--Higgs measurement obtains its sensitivity.

The presence of an additional light scalar in such an enhanced-coupling scenario is not at all exotic.  Indeed, one or more relatively light new scalars are \emph{required} in weakly-coupled extended Higgs sectors when the couplings of $h$ are significantly modified, due to the decoupling behavior of these theories~\cite{HaberDecoupling}.  For example, in the Georgi-Machacek model, a simultaneous enhancement of the $h$ couplings to fermions and vector bosons by more than 10--15\% can only be obtained when at least one of the new scalars lies below about 400~GeV~\cite{Hartling:2014aga}.
Such a relatively light additional scalar resonance can hide from direct searches if it decays predominantly into the same undetected nonstandard final states that give rise to ${\rm BR}_{\rm new}$ of $h$.

By contrast, when $H$ is heavy, $m_h^2 \ll p^2 < m_H^2$, the scalar-exchange amplitude becomes
\begin{equation}
	\mathcal{\bar M} = 1 + \Delta \frac{m_H^2 - m_h^2}{m_H^2 - p^2}
	\simeq (1 + \Delta) + \Delta \frac{p^2}{m_H^2}.
\end{equation}
This reduces to the situation in which only $h$ is exchanged, $\mathcal{\bar M} = 1 + \Delta$, in the limit $m_H^2 \gg p^2$.  As $p^2$ approaches $m_H^2$ from below, the presence of the $H$ resonance manifests as a momentum-dependent rise in the amplitude, which can be reinterpreted in terms of a dimension-six operator obtained by integrating $H$ out of the theory.

%%%%%%%%%%%%%%%%%%%%%%%%%%%%%%%%%%%%%%%%%%%%%%
%\section{off-shell Higgs production in vector boson fusion}

Finally I comment on the effect of light unitarizing scalars on off-shell Higgs production via vector boson fusion, $WW \to h^* \to ZZ$.  This channel is included in the CMS off-shell--Higgs analysis~\cite{Khachatryan:2014iha}, though its contribution to the sensitivity is currently very small due to the small event rate.  This channel has the benefit of avoiding model dependence due to new light colored degrees of freedom that would affect the $ggh$ coupling.  Its sensitivity, however, still relies on the fact that when $\kappa_W^h \kappa_Z^h \neq 1$, the discovered Higgs boson $h$ by itself no longer unitarizes the $WW \to ZZ$ scattering amplitude at high energy~\cite{Lee:1977yc}, leading to quadratic growth of this amplitude with increasing center-of-mass energy.  In models with an extended Higgs sector that yield $\kappa_{W,Z}^h > 1$, unitarity of the $WW \to ZZ$ scattering amplitude is restored by additional diagrams involving $t$-channel exchange of a singly-charged Higgs boson that couples to $W^+Z$~\cite{Falkowski:2012vh,Grinstein:2013fia,Bellazzini:2014waa}, yielding the sum rule
\begin{equation}
	\sum_i \kappa_W^{h_i} \kappa_Z^{h_i} = 1 + (\kappa_{WZ}^{H^+})^2,
\end{equation}
where the sum runs over the neutral states $h$, $H$, and any others that couple to $W$ and $Z$ pairs, and the $H^+W^-_{\mu}Z_{\nu}$ Feynman rule is $2i M_W M_Z \kappa_{WZ}^{H^+} g_{\mu\nu} / v$.
The related process $WW \to WW$ is unitarized by a doubly-charged Higgs boson with a coupling to like-sign $W$ pairs, yielding the sum rule~\cite{Falkowski:2012vh,Grinstein:2013fia,Bellazzini:2014waa}
\begin{equation}
	\sum_i (\kappa_W^{h_i})^2 = 1 + (\kappa_W^{H^{++}})^2,
	\label{eq:Hppsumrule}
\end{equation}
where again the sum runs over all neutral states that couple to $WW$ and the $H^{++}W^-_{\mu} W^-_{\nu}$ Feynman rule is $2i M_W^2 \kappa_W^{H^{++}} g_{\mu\nu}/v$.  These states appear in the Georgi-Machacek model and its generalizations, as well as in the septet model.
As in the gluon-fusion case, if the additional unitarizing Higgs boson(s) are all light compared to the $m_{ZZ}$ (or $m_{WW}$) range in which the LHC measurements would obtain their sensitivity, the growth of the $h$-exchange amplitude with increasing $m_{ZZ}$ is largely cancelled and the residual deviation of the amplitude from its SM value again falls like $\mathcal{\bar M} - 1 \propto m_H^2/p^2$.

A light singly- or doubly-charged Higgs boson is more difficult to hide from direct searches than a second neutral Higgs boson, because it cannot generically decay into the same undetected light new physics that gives rise to ${\rm BR}_{\rm new}$ of $h$.  A dedicated search for a fermiophobic singly-charged Higgs boson produced in $WZ$ fusion and decaying to $WZ$ was recently performed by ATLAS~\cite{Aad:2015nfa}; however, such a charged Higgs with mass between half the $Z$ mass and 200~GeV remains largely unconstrained by direct searches.  For the doubly-charged Higgs, LHC measurements of the like-sign $W^{\pm}W^{\pm}$ cross section already put rather stringent constraints on production of a doubly-charged Higgs boson in vector boson fusion if it decays solely to $W^{\pm}W^{\pm}$~\cite{Chiang:2014bia}.  The unitarity sum rule in Eq.~(\ref{eq:Hppsumrule}) then allows this direct-search limit to be translated into a model-independent upper bound on $\kappa_{W,Z}^h$~\cite{Logan:2015xpa}, as a function of the doubly-charged Higgs boson's mass.  The constraints are least stringent for a doubly-charged Higgs below 100~GeV, where the existing LHC measurement loses sensitivity due to the increasingly soft charged leptons from the off-shell $W$ decays~\cite{Chiang:2014bia}.

%%%%%%%%%%%%%%%%%%%%%%%%%%%%%%%%%%%%%%%%%%%%%%
\section{Summary}
\label{sec:conclusions}

In this paper I showed that the interpretation of LHC measurements of the $gg \ (\to h^*) \to ZZ$ cross section in the high $m_{ZZ}$ region as a constraint on the Higgs total width is invalidated if new light scalar degree(s) of freedom that unitarize the $t \bar t \to ZZ$ scattering amplitude are present at energy scales below that at which the LHC measurement obtains its sensitivity.  In particular, the $gg \to ZZ$ cross section in the high $m_{ZZ}$ region can be very SM-like even in the case that the product of couplings $\kappa_t^h \kappa_Z^h$ of the 125~GeV Higgs boson $h$ is substantially larger than predicted in the SM if a second light neutral Higgs boson $H$ with appropriate couplings is present.  A similar conclusion follows for the process $WW \ (\to h^*) \to ZZ, WW$ so long as the unitarity-restoring singly- or doubly-charged Higgs boson is light.

%%%%%%%%%%%%%%%%%%%%%%%%%%%%%%%%%%%%%%%%%%%%%%
\begin{acknowledgments}
I thank J.~Gunion, M.-J.~Harris, K.~Kumar, V.~Rentala, and M.~Spannowsky for thought-provoking conversations and S.~Godfrey for comments on the manuscript.
This work was supported by the Natural Sciences and Engineering Research Council of Canada.  \end{acknowledgments}
%%%%%%%%%%%%%%%%%%%%%%%%%%%%%%%%%%%%%%%%%%%%%%

%\appendix

%%%%%%%%%%%%%%%%%%%%%%%%%%%%%%%%%%%%%%%%%%%%%

\end{document}